\documentclass[twocolumn,showpacs,preprintnumbers,amsmath,amssymb]{revtex4}
\usepackage{graphicx}
\usepackage{dcolumn}
\usepackage{bm}

\newcommand{\be}{\begin{equation}}
\newcommand{\ee}{\end{equation}}
\newcommand{\bea}{\begin{eqnarray}}
\newcommand{\eea}{\end{eqnarray}}

\newcommand{\nn}{\nonumber}

\newcommand{\ov}{\overline}

\newcommand{\un}{{\bf 1}}

\newcommand{\f}{{\bf 5}}

\newcommand{\te}{{\bf 10}}

\newcommand{\op}{\oplus}

\newcommand{\beq}{\begin{equation}}
\newcommand{\eeq}{\end{equation}}

\newcommand{\cF}{\mathcal{F}}

\newcommand{\cV}{\mathcal{V}}

\newcommand{\R}{\text{Re}}

\newcommand{\tv}{\text{v}}

\newcommand{\cref}{{\bf [check ref]}}

\begin{document}
\preprint{MPP-2011-53}
\preprint{CPHT-RR040.0511}
\title{On Fluxed Instantons and Moduli Stabilisation in IIB Orientifolds and F-theory
}
\author{Thomas W.~Grimm$^{1}$, 
             Max Kerstan$^{2}$, 
             Eran Palti$^{3}$, 
             Timo Weigand$^{2}$}
\affiliation{%
$^{1}$ Max-Planck-Institut f\"ur Physik, 
               Munich, Germany 
\\
$^{2}$ Institut f\"ur Theoretische Physik, Ruprecht-Karls-Universit\"at Heidelberg,
             Germany
\\
$^{3}$ Centre de Physique Theorique, Ecole Polytechnique,
               CNRS, 
               Palaiseau, France
}
\begin{abstract}
We study the superpotential induced by Euclidean D3-brane instantons  
carrying instanton flux, with special emphasis on its significance for 
the stabilisation of K\"ahler moduli and Neveu-Schwarz axions in Type IIB orientifolds. 
Quite generally, once a chiral observable sector is included in the  
compactification, arising on intersecting D7-branes with world-volume  
flux, resulting charged instanton zero modes prevent a class of  
instantons from participating in moduli stabilisation. We show that 
instanton flux on Euclidean D3-branes can remove these extra zero  
modes and helps in reinstating full moduli stabilisation within a  
geometric regime. We comment also on the F-theoretic description of  
this effect of alleviating  the general tension between moduli  
stabilisation and chirality. In addition we propose an alternative  
solution to this problem based on dressing the instantons with charged  
matter fields which is unique to F-theory and cannot be realised in  
the weak coupling limit.
\end{abstract}
\maketitle

\section{Introduction and Summary}

Moduli stabilisation is essential to determine the vacuum structure of string compactifications. Despite progress in other corners of the string landscape, moduli stabilisation is currently best under control in Type IIB string theory compactifications.  Following \cite{Giddings:2001yu}  the complex-structure moduli are fixed perturbatively by closed-string fluxes, while in the setup of \cite{Kachru:2003aw} the K\"ahler moduli are fixed non-perturbatively, either through gaugino condensation on D7-branes or through Euclidean D3-instantons, called E3-instantons in the sequel (see, however, \cite{Balasubramanian:2005zx} for other proposals). While this is the picture for the hidden closed-string sector, the visible open-string sector should arise from intersecting D7-branes with world-volume flux or from D3-branes on singularities \cite{Aldazabal:2000sa}. In this paper we will study the former possibility and work in the geometric regime where the volumes of all cycles are larger than the string scale and so $\alpha'$ corrections are under control. 

Consider, for simplicity, an $SU(5)$ GUT model realised by a stack of D7-branes wrapping a Calabi-Yau divisor $[D_A]$ and a single D7 wrapping a divisor $[D_B]$. The resulting gauge group is $SU(5)\times U(1)_A\times U(1)_B$. To realise a chiral sector in the $\bf{10}$ representation a gauge flux ${\cal F}_A$ must be turned on along the $A$ stack along the diagonal $U(1)_A$ and supported on the two-cycle intersection between the divisor $[D_A]$ and the orientifold locus $D_A \cap D_{O7}$. Consider now an E3-instanton wrapping a divisor $[D_E]$ which only intersects $[D_A]$ along the locus where the flux ${\cal F}_A$ is turned on. In this case there will be open-string zero-modes between the D7-brane and the E3-instanton which are charged under $U(1)_A$ \cite{Blumenhagen:2006xt}. The presence of such modes implies that a possible contribution of such an instanton to the superpotential necessarily involves charged open string fields. 
In \cite{Blumenhagen:2007sm} it was realised that this implies a certain tension between a chiral open-string sector and stabilising the moduli non-perturbatively. Namely, as we will recall momentarily, only such instantons can be relevant for moduli stabilisation 
whose zero modes do not carry net $U(1)_A$ charges. In general these instantons will not suffice to fix all K\"ahler moduli non-perturbatively.

It is informative to consider this situation from a macroscopic perspective. The effect of nonzero D7 flux supported on the intersection $D_A \cap D_E$ of brane and E3-instanton is to gauge the axionic component of the chiral superfield $T_E$ associated to the divisor $D_E$. This implies that the instantonic exponential $e^{-T_E}$ is not gauge invariant by itself \cite{Blumenhagen:2006xt,Haack:2006cy}. Rather, the instanton contribution to the superpotential must be dressed with open-string modes $\Phi_i$ that are charged under $U(1)_A$ to cancel the gauge transformation. Hence the contribution takes the schematic form
\be
W \supset \left(\prod_i \Phi_i \right) e^{-T_E} \;.
\ee
The problem pointed out in \cite{Blumenhagen:2007sm} is that since the fields $\Phi_i$ are charged under the $U(1)_A$ they must also be charged under the $SU(5)$. This means that they should not develop a VEV else they would break the visible gauge group \footnote{Note that the same problem occurs for the non-abelian parts of the Standard Model gauge group in non-GUT realisations of the MSSM.}. Hence the corresponding superpotential contribution does not participate in closed string moduli stabilisation.

In this article we study moduli stabilisation in Type IIB orientifold setups with 2-forms of negative parity under the holomorphic involution \cite{Grimm:2004uq}. In particular we are concerned with the consequences of the presence of odd forms on the above discussion regarding the gauge invariance of instantons. Such forms lead to additional supergravity fields that also appear in the instanton action and which transform under the gauge transformation. These complex moduli from the R-R and NS-NS two-forms have been 
argued to correct the E3-brane superpotential in the absence of D7-branes via a Jacobi theta function summing over all lower-dimensional brane charges \cite{Grimm:2007xm}, following works on M5-brane instantons \cite{Witten:1996bn}.
We will show that, even in the presence of an intersecting D7-brane, gauge invariant contributions will remain to contribute to the 
superpotential and the moduli stabilisation problem is alleviated. A key aspect of this result is that we should sum over all instantonic contributions including instanton configurations which support world-volume flux. Establishing a superpotential contribution of fluxed instantons requires a detailed analysis of the instanton consistency conditions, which is another purpose of this article. It turns out that in the case of $O(1)$-instantons the orientifold projection allows only negative parity flux. We find that only instanton fluxes which can be written as the pullback of non-trivial forms in the orientifold bulk contribute to the total charge of the instanton. Hence instanton fluxes do not affect the selection rules of $O(1)$-instantons in orientifold setups without negative parity 2-forms, which was the initial setup considered in \cite{Blumenhagen:2007sm}.

Consequently in a setup with negative parity 2-forms some fluxed instanton configurations are gauge invariant under $U(1)_A$ even though an unfluxed instanton along the same divisor would be charged. These fluxed instantons can then contribute to the superpotential by themselves without the inclusion of open-string modes. Microscopically such configurations do not have any \emph{net} chiral charged zero modes. The fluxed instantons can therefore provide the additional superpotential contributions required to fix all moduli fields in a supersymmetric vacuum in a geometric regime. As we will discuss, vector-like charged zero modes may still prevent a superpotential contribution, but this must be analysed in concrete examples. 

As we briefly discuss, analogous selection rules apply for M5-instantons in F-theory compactifications which admit a Type IIB weak coupling limit. In particular we discuss subtleties of the uplift arising if the $U(1)$ in the Type IIB weak coupling limit acquires a mass by the \emph{geometric} St\"uckelberg mechanism, i.e. independently of possible gauge flux. Such $U(1)$ symmetries can be described consistently by an expansion of the M-theory three-form $C_3$ into non-harmonic forms \cite{Grimm:2010ez, appear}. If no fluxes are turned on along this geometrically massive $U(1)$ it can be integrated out consistently at the Kaluza-Klein scale. The selection rules imposed upon M5-instantons by these gauge symmetries remain as accidental symmetries in the low-energy theory. Thus the same instanton selection rules apply as in the Type IIB weak coupling limit despite the fact that the responsible $U(1)$ symmetry is not directly visible below the KK-scale.

We also observe that F-theory allows for a further way to circumvent the tension between moduli stabilisation and realising a chiral matter sector which has no analogue in the Type IIB weak coupling limit. The key ingredient intrinsic to F-theory is the possible local enhancement to an exceptional gauge group. Decomposing the adjoint of these exceptional groups yields modes which are charged under the $U(1)$ on which the chirality-inducing flux is turned on but which are singlets under the visible gauge group. These modes can therefore acquire a non-zero VEV without leading to phenomenological problems. This implies that even instanton configurations which are not gauge invariant by themselves can contribute to the superpotential when dressed with a suitable combination of such charged open-string fields.

This article is organised as follows: In section \ref{sec:macu1} we describe the fluxed instanton configurations from a macroscopic perspective using the gauged supergravity formalism. In section \ref{sec:micu1} we investigate these instantons from a microscopic perspective, demonstrating the absence of net extra charged zero modes for suitable instanton flux and also discussing the neutral zero modes. In section \ref{sec:modst} we study the effect of fluxed instantons on moduli stabilisation and show that, at least in principle, all K\"ahler moduli can be stabilised in a supersymmetric vacuum in a geometric regime. In section \ref{sec:instf} we briefly outline how such configurations are lifted to F-theory and also discuss the second proposed mechanism of giving singlet fields a vev to neutralise the instanton. 

The expressions in this article are very general and serve as a candidate solution to a general problem. It would be interesting to see how the ideas presented are realised in an explicit Calabi-Yau setup within a global orientifold compactification with a chiral matter sector e.g. of the type constructed in \cite{Blumenhagen:2008zz}. Even within a general setup the implications of the moduli stabilisation mechanism can be important for phenomenological aspects (for example soft masses \cite{Blumenhagen:2009gk}). It would be very interesting to study the phenomenology resulting from the proposed moduli stabilisation framework.

\section{Macroscopics: $U(1)$ charges}
\label{sec:macu1}

\subsection{Supergravity background}

We begin by reviewing aspects of compactifications of Type IIB string theory on a Calabi-Yau threefold $X_3$, modded out by the orientifold action $\Omega (-1)^{F_L} \sigma$ \cite{Grimm:2004uq}. Under the induced geometric action of $\sigma$ the cohomology groups $H^{(p,q)}(X_3)$ split into $H^{(p,q)}_{\pm}(X_3)$. Our notation for a basis of the two- and four-forms is summarised in table I.
\begin{table}[htbp] 
\renewcommand{\arraystretch}{1.5} 
\begin{center} 
\begin{tabular}{|c|c|c|} 
  \hline 
  cohomology group & basis& index range\\ 
  \hline \hline 
$H^{(1,1)}_+$ &   $ \omega_{\alpha}$   & $\alpha= 1, \ldots h_+^{(1,1)}$ \\ 
$H^{(1,1)}_-$ &   $ \omega_{a}$   & $a= 1, \ldots h_-^{(1,1)}$ \\ 
$H^{(2,2)}_+$ &   $ \tilde \omega^{\alpha}$   & $\alpha= 1, \ldots h_+^{(1,1)}$ \\ 
$H^{(2,2)}_-$ &   $ \tilde \omega^{a}$   & $a= 1, \ldots h_-^{(1,1)}$ \\ 
\hline
\end{tabular} 
\label{cohom1}
\caption{Even cohomology expansion basis.} 
\end{center} 
\end{table}
Since the volume form is even under the involution the non-trivial intersection numbers are given by
\bea
\kappa_{\alpha \beta \gamma} &=&  \int _{X_3}  w_{\alpha} \wedge  w_{\beta} \wedge w_{\gamma}, \nn \\
\kappa_{\alpha b c} &=& \int _{X_3}  w_{\alpha} \wedge  w_{b} \wedge w_{c} \;.
\eea
The two- and four-forms in table I are dual in the sense that
\bea
\int_{X_3} \omega_a \wedge \tilde \omega^{b} = \delta_a^b, \quad\quad \quad \int_{X_3} \omega_{\alpha} \wedge \tilde \omega^{\beta} =  \delta_{\alpha}^{\beta} \;. 
\eea
The K\"ahler form $J$ of $X_3$ and the R-R and NS-NS forms have an expansion 
\bea
J &=& v^{\alpha} \omega_{\alpha}, \quad\quad  C_2 = c^{a} \omega_a \;, \\
B_2 &\equiv&   B_- + B_+ = b^{a} \omega_a + b^{\alpha} \omega_{\alpha}\;, \\
C_4 &=& c_{\alpha} \tilde \omega^{\alpha} + c_2^{\alpha} \wedge \omega_{\alpha} + \ldots \, ,
\eea
where we restricted ourselves to expansion along the even-dimensional cohomology. Note that in the above the component of the $B$-field $B_+$ along the involution even cycles, $b^{\alpha}$, is not a continuous modulus but can only take the discrete values $0, \frac{1}{2}$ consistent with the orientifold action.

We use the conventions of \cite{Blumenhagen:2006ci} for the signs of the kinetic terms of the bulk moduli as well as the Chern-Simons action of the D7-branes. The appropriate chiral fields for these compactifications are given by \cite{Grimm:2004uq,Jockers:2004yj}
\bea
G^a &=& c^{a} - \tau b^a,  \label{GaTalpha} \\
T_{\alpha} &=& \frac{1}{2}\kappa_{\alpha \beta \gamma} v^\beta v^\gamma+ i \left(c_{\alpha} - \kappa_{\alpha bc} c^b {b}^c \right) + \frac{i}{2} \tau \kappa_{\alpha b c}  b^b   \, b^c \;, \nonumber
\eea
where $\tau = C_0 + i\, e^{-\phi}$ represents the axio-dilaton. 

We consider now a stack of $N_A$ D7-branes  along the holomorphic divisor $D_A$ together with its orientifold image $D_{A'}$.
It is convenient to define the objects
\bea \label{def-D+D-}
D_A^+ = D_A \cup D_{A'}, \quad\quad\quad D_A^- = D_A \cup ( - D_{A'}) \;,
\eea
with Poincar\'e dual classes $[D^{\pm}_A]$. Here $- D_{A'}$  is orientation reversed with respect to $D_{A'}$.
The corresponding wrapping numbers along the basis elements of $H_4^{\pm}(X_3, \mathbb Z)$ are
\bea
C^{\alpha}_A = N_A \int_{D_A^{+}} \tilde \omega^{\alpha}, \quad\quad  
C^{a}_A = N_A  \int_{D_A^{-}} \tilde \omega^a, 
\eea
where we have included an otherwise omnipresent factor of $N_A$ for notational convenience.

If $D_A \neq D_{A'}$, this stack of $N_A$ D-branes gives rise to a $U(N_A)$ gauge theory. We will concentrate on its diagonal $U(1)$ gauge factor in the sequel. 

The gauge field $\hat{F}_A $ splits into the field strength  $F_{A}$ in four dimensions and the internal gauge flux ${\cal F}_{A}$ along the cycle wrapped by the divisor. The latter receives two types of contributions: flux that lies in the image of the pullback $\i^*$ of two-forms from the bulk onto the brane and 
flux in the complement of $\i^* H^2(X_3)$. On divisors of Calabi-Yau three-folds it is possible (though not necessarily in a unique manner) to choose a basis of these two subspaces  that is mutually orthogonal with respect to the wedge product \cite{Jockers:2004yj}. Flux on $D_A$ expressed in terms of such basis elements of the compliment of $\i^*$ will be called variable flux ${\cal F}_A^{\tv}$.

The gauge flux ${\cal F}_{A}$ on $D_A$ can then be expanded as 
\bea
2 \pi \alpha' {\cal F}_{A} =  {\cal F}^{a}_{A} \omega_a  +   {\cal F}^{\alpha}_{A} \omega_{\alpha}  + {\cal F}_A^{\tv},
\eea
where we are suppressing the explicit pullback in the first two terms.\footnote{The gauge field has dimension of inverse length squared, so that $2\pi\alpha' \cF_A$ is dimensionless. All other fields as well as the basis forms in \eqref{cohom1} are chosen dimensionless.} We also refrain from expanding  the variable flux $ {\cal F}_A^{\tv}$ into an explicit basis since this will play no role in our discussion.

Finally since the field strength $\hat{F}_A $ appears in the Chern-Simons and DBI action only in the gauge invariant combination $2 \pi \alpha'\hat{F}_{A}  -  \i^* B$ it is convenient to combine the pullback gauge flux and the background $B$-field into the quantity
\bea
\label{frakfdef}
\tilde{\cal F}_A = 2 \pi \alpha'{\cal F}_{A}  -  \i^* B
\eea
with analogous components along the 2-cohomology.

In the presence of open-string fluxes of the pullback type some of the closed string axions become charged under the open-string $U(1)$s. For the purposes of our analysis we wish to extract this gauging with respect to the diagonal $U(1)$ taking into account the transformation of the orientifold odd fields. The relevant terms follow by dimensional reduction of the Chern-Simons coupling to linear order in the four-dimensional field strength $F_{A}$. To this end we  take into account both the contributions from the brane along $D_A$ and its orientifold image along $D_{A'}$ and divide the result by $2$ to arrive at the physical action defined on the orientifold quotient. The resulting gauging reads
\bea
\nabla G^a &=& d G^a - Q_A^a A^A,  \label{gauging1} \\
\nabla T_\alpha &=& dT_\alpha - i  Q_{A \alpha} A^A \;, \label{gauging2}
\eea 
where we have introduced the Killing vectors
\bea
\label{St-coupl}
Q_A^a &=&    2\pi\alpha'  C^a_A \;, \label{Qsodd} \\
Q_{A \alpha} &=&    -  2\pi\alpha' \Bigl( \kappa_{\alpha \beta \gamma}  \, \tilde{\cal F}_A^{\beta} \, C^{\gamma}_A  +   \kappa_{\alpha b c }  \, {\cal F}_A^b   \, C^{c}_A \Bigr) \;. \label{Qseven}
\eea
Note that the gauging of the $G^a$ is independent of any gauge flux and leads to a geometric St\"uckelberg mechanism, while the gauging of the $T_\alpha$ is entirely flux-induced.
The gauging of the axions will allow us to determine the transformation properties of the instanton superpotential contribution, to which we now turn.

\subsection{Fluxed instantons in supergravity}
\label{sec:flux_Inst}

We are interested in the  superpotential contribution of an E3-instanton wrapping a holomorphic divisor.
As for spacetime-filling 7-branes, the instanton effective action  is defined in terms of the combinations $D_E^\pm = D_E \cup (\pm D_{E'})$. 
Our interest is in the effect of two crucial ingredients: worldvolume flux ${\cal F}_E$ on the instanton and the presence of orientifold-odd moduli. 
In the next section we will discuss the microscopic selection rules determining which instanton fluxes can in principle yield a superpotential contribution. For example, for a rigid instanton along a cycle $D_E = D_{E'}$, not pointwise invariant, the allowed fluxes are all elements in the lattice $H^{1,1}_-(D_E,\mathbb Z)$.
In the following expressions we would like to be more general concerning the properties of $D_E$ but always assume that only fluxes from the allowed flux lattice are being considered.

We first  ignore the gauging of the orientifold odd moduli induced by spacetime-filling 7-branes.
The superpotential contribution of a single E3-instanton along $D_E^+$ then involves a summation over all allowed pullback and variable instanton fluxes,
\bea
W \simeq  \sum_{{\cal F}_E}  {\rm exp} (- S_{E}), \quad\quad\quad  S_{E} =   (2\pi\alpha')^{-2} {f}_E.   
\eea
 Here ${f}_E =  {f}_E(T_\alpha, G^a; {\cal F}_E)$ denotes the gauge kinetic function
associated with a hypothetical spacetime-filling D7-brane wrapped along the same internal divisor. It depends manifestly on the K\"ahler chiral multiplet and the $G^a$-moduli and also on the internal gauge flux. The gauge kinetic function $f_E$ can be determined by dimensional reduction of the DBI and CS action of a spacetime-filling 7-brane along $D_E$ to second order in the four-dimensional field strength. It takes the form \footnote{The gauge kinetic function is defined with respect to the action $S = \int_{\mathbb R^{1,3}}  \frac{1}{2} {\rm Re}( f) F \wedge \ast F + \frac{1}{2} {\rm Im} (f) F \wedge F$ and includes again a factor of $\frac{1}{2}$ due to the ${\mathbb Z}_2$ projection.} \footnote{Note that the universal contribution of the $b^a$-moduli is already encapsulated in $T_\alpha$ and $G^a$ given in (\ref{GaTalpha}) so that only the actual pullback gauge flux  ${\cal F}^a_E$ (as opposed to $\tilde {\cal F}^a_E$ ) appears in $\Delta_{E \alpha}$ and $\Delta_{E a}$.}
\bea
\label{gaugekina}
 {f}_E   &=&  \pi (2\pi\alpha')^2 \Big( C^\alpha_E ( T_{\alpha} + i  \Delta_{E \alpha}) 
                     +  i C^a_E  \Delta_{E a} +    i  \Delta^{\tv}_{E}      \Big), \nonumber \\
\Delta_{E \alpha} &=&    \kappa_{\alpha b c} \, G^b \, { \cal F}^c_E  + \frac{\tau}{2} \Big( \kappa_{\alpha b c}   { \cal F}^b_E \,   { \cal F}^c_E         
 + \kappa_{\alpha \beta \gamma}  {\tilde{ \cal F}}^\beta_E  \,  {\tilde{ \cal F}}^\gamma_E \Big),  \nonumber \\
\Delta_{E a} &=& \kappa_{a b \gamma} G^b \, {\tilde{ \cal F}}_E^\gamma +    \tau \, \kappa_{a b \gamma} \,   { \cal F}^b_E  \, {\tilde{ \cal F}}_E^\gamma, \\
\Delta_{E}^{\tv} &=&  \tau \, \int_{D_E} {\cal F}_E^\tv \wedge  {\cal F}_E^\tv  \nn .
\eea

Performing the sum over admissible instanton gauge fluxes leads to the appearance of theta-functions in the instanton partition function depending in particular on the 
VEV of the odd moduli $G^a$ \cite{Grimm:2007xm}. 
This parallels the situation for the partition function of M5-instantons in M/F-theory \cite{Witten:1996bn}.
For special values of the $G^a$-moduli corresponding to the zeros of the theta-functions the E3-brane superpotential vanishes.
This is immediately clear if one considers that the summation over fluxes can lead to cancellations from the complex phases in ${\rm exp}(-S_E)$, which depend both on the fluxes and the VEV of the $G^a$-moduli. 
The phenomenon that the summation of individually non-zero terms may lead to cancellations is of course familiar e.g. from the context of heterotic worldsheet instantons \cite{Silverstein:1995re}.

We now proceed to include the gauging of the axions $c^a$ and $c_\alpha$ given in eq. (\ref{gauging1}) and (\ref{gauging2}) induced by a spacetime-filling D7-brane along a different divisor $D_A$.
Even before performing the sum over instanton fluxes, we note that  for each individual instanton flux configuration appearing in the instanton partition function
the classical instanton effective action $S_{E} $ itself shifts non-trivially under a gauge symmetry $U(1)_A$. 
As can be deduced from the expression (\ref{gaugekina}) this instanton flux dependent gauge shift is of the form
\bea
\label{U1-trafo}
 A^A &\rightarrow& A^A + \frac{1}{\ell_s^2} d \Lambda^A, \quad e^{-S_E} \rightarrow e^{i q_A \Lambda^A} e^{-S_E},   \nn \\
q_A &=&  \frac{1}{2} \Big(  \, \kappa_{\alpha b c } \, C^{\alpha}_E  \, C^b_A \, ({\cal F}^c_A - {\cal F}^c_E) + \kappa_{\alpha \beta \gamma} C^\alpha_E \, C^\beta_A \, {\tilde{\cal F}}^\gamma_A  \nn \\
&& - \kappa_{a b \gamma} C^a_E C^b_A {\tilde{\cal F}}_E^\gamma \Big).
\eea
Note that only the pullback instanton flux contributes to the instanton charge, not the variable flux ${\cal F}_E^\tv$. 

As a consequence of the non-trivial $U(1)_A$ charge $q_A$ associated with a given instanton flux, this configuration
contributes now to a superpotential of the form $W \simeq {\cal O} \, {\rm exp}(-S_E)$, where  ${\cal O} $ is an operator involving open string fields with $U(1)_A$  charge $- q_A$.
The sum over the instanton fluxes in the instanton partition function then splits into various contributions. Each such contribution involves the sum over the full allowed lattice of variable instanton fluxes ${\cal F}_E^\tv$ together with the sublattice of allowed pullback fluxes  whose induced charge matches that of the respective operator ${\cal O}$.

In particular, for the factor ${\rm exp}(-S_E)$ to be gauge invariant by itself the charge $q_A$ must vanish. The analysis of  \cite{Blumenhagen:2007sm} restricts to the second term on the right-hand-side of (\ref{U1-trafo}), which is the only term present in the absence of orientifold odd cycles. However we see that in the more generic setting with $h^{1,1}_-(X_3) > 0$  there are additional contributions. These additional contributions can cancel each other leaving the instanton gauge invariant. The sum over instanton fluxes ensures that the presence of such a gauge invariant combination is generic. This is in contrast to the case with only orientifold even cycles where the possibility of the second term of (\ref{U1-trafo}) vanishing by itself requires tuning of the gauge flux ${\cal F}_A$.

In the next section we show that, as expected, neutrality under $U(1)$ charges is equivalent to absence of net chiral charged zero modes. 

\section{Microscopics: zero modes}
\label{sec:micu1}

We now turn to a microscopic description of fluxed E3-instantons, addressing in particular the effect of instanton fluxes on both the uncharged and charged zero modes.
If the holomorphic divisor wrapped by the Euclidean D3-brane is not invariant under the orientifold, $D_E \neq D_{E'}$, the instanton is conventionally called a $U(1)$ instanton, while if $D_E = D_{E'}$ without being pointwise invariant (i.e. not on top of the O7-plane) one speaks of an $O(1)$ instanton \footnote{The situation $D_E = D_{E'}$ pointwise will not be considered here as such instantons do not contribute to the superpotential in orientifolds with standard O7-planes.}.
  Neutral instanton zero modes arise from open string excitations with both ends on an instanton and fall into the following three classes:

The \emph{universal} zero modes are the Goldstone modes associated with the breakdown of four-dimensional super-Poincar\'e invariance due to the localisation of the instanton in spacetime.
For a general BPS instanton these are the four bosonic position modes $x^{\mu}$ and their superpartners $\theta^{\alpha}$ and $\ov \tau^{\dot \alpha}$. Effectively $x^{\mu}$ and $\theta^{\alpha}$ can be identified as entering the superpotential measure $d^4 x \, d^2 \theta$. By contrast, $\ov \tau^{\dot \alpha}$ are Goldstinos associated with the  ${\cal N}=1$ subalgebra of the original ${\cal N}=2$ supersymmetry  which is not preserved by the orientifold action. 
The \emph{geometric} zero modes include the deformation modes of the instanton divisor as well as potential Wilson line moduli.
\emph{Localised} neutral zero modes can arise at the intersection of a $U(1)$ instanton along $D_E$ with its orientifold image $D_{E'}$. 

An instanton contributes to the superpotential if all neutral zero modes other than the universal $x^{\mu}, \theta^{\alpha}$ are either projected out by the orientifold action or saturated by suitable non-derivative couplings in the instanton effective action. 
For $O(1)$ instantons the $\overline \tau^{\dot \alpha}$-modes are projected out by the orientifold action \cite{Argurio:2007qk}. Therefore, if the divisor $D_E$ has no deformation and Wilson line moduli, an $O(1)$ instanton straightforwardly contributes to the superpotential. 
For $U(1)$ instantons this is not the case and the saturation of the universal $\ov \tau^{\dot \alpha}$-modes is notoriously difficult. Different mechanisms involving various types of couplings in the instanton effective action have been identified in the literature \cite{Blumenhagen:2007bn}. 

Our prime interest in this note is in $O(1)$ instantons. An immediate question is whether in the presence of instanton flux the universal $\overline \tau^{\dot \alpha}$-modes continue to be projected out such that the instanton can contribute to the superpotential.
To appreciate the nature of $O(1)$ instantons we nonetheless begin with a description of flux along a $U(1)$ instanton $D_E$ and its image $D_{E'}$.
Under the orientifold action the gauge invariant combination of instanton flux and $B$-field background value $\tilde {\cal F}_E$ along $D_E$ is mapped to
 $\tilde {\cal F}_{E'} = - \sigma^* \tilde {\cal F}_{E}$ along $D_{E'}$.  This is because the orientifold acts in the same manner on all purely \emph{internal} instanton fields as it would for the modes of a  D7-brane wrapping the same divisor.

In order for the fluxed instanton $(D_E, \tilde {\cal F}_E)$ to contribute to the superpotential, it must be half-BPS. 
This subjects the flux to a D-term and F-term  supersymmetry condition. The D-term reads
\bea
\label{DSUSY1}
\int _{D_E} J \wedge \tilde {\cal F}_E  = \frac12  \int_{D_E^+} J \wedge \tilde {\cal F}_E^+   +   \frac12  \int_{D_E^-} J \wedge \tilde {\cal F}_E^- =  0,
\eea
 which is the direct analogue of the D-term supersymmetry condition for D7-brane fluxes. Note that the D-term depends only on the instanton flux arising by pullback from the ambient space as for variable flux ${\cal F}_E^\tv$ the integral vanishes identically.
 The F-term supersymmetry condition on $\tilde {\cal F}_E$ amounts to requiring that $\tilde {\cal F}_E \in H^{1,1}(D_E)$. This is trivially satisfied for rigid instantons, for which $H^{2,0}(D_E) = 0$.

 Now consider an $O(1)$ instanton instead. Our claim is that if $D_E$ is rigid, the configuration $(D_E, \tilde {\cal F}_E^-)$ continues to contribute to the superpotential provided the instanton flux is purely odd, i.e. of type $\tilde {\cal F}_E^- \in H^{1,1}_-(D_E)$ only.
 In fact, the instanton plus flux is invariant as a whole under the orientifold involution. Therefore the projection of the $\ov \tau^{\dot \alpha}$-modes is unchanged compared to unfluxed $O(1)$ instantons. 
 An important property of this configuration is that the D-term supersymmetry condition on the instanton flux (\ref{DSUSY1}) is automatically satisfied for every choice of K\"ahler form:  For an $O(1)$ instanton $D_E^- = 0$ and as stressed above we restrict ourselves to $\tilde {\cal F}^E_+ =0$. 
 Note that a non-trivial D-term condition would imply the appearance of lines of marginal or threshold stability from the inclusion of gauge flux. This in turn would be in conflict with holomorphicity of the superpotential \cite{GarciaEtxebarria:2007zv}. Consistently with these general considerations $O(1)$ instantons with odd flux are half-BPS (in the large volume regime) with respect to every choice of K\"ahler moduli.

 However, an extra constraint arises from the Freed-Witten quantisation condition on the gauge flux. As for spacetime-filling branes, the quantisation condition is
 \bea
 {\tilde {\cal F}}_E + \i^* B + \frac{1}{2}c_1(K_{D_E}) \in H^2(D_E, \mathbb Z).
 \label{FWanom}
 \eea
 For an $O(1)$ instanton this has the following consequences: Since  $[D_E] = \frac12 [D_E^+]$, the canonical bundle satisfies  $c_1(K_{D_E}) \in H^2_+(D_E)$. Thus the orientifold even part of the $B$-field must cancel the half-integer contribution for a non-spin divisor $D_E^+$ because, as established above,  the gauge flux  must be odd under the orientifold action, $\tilde {\cal F}_E^+=0$.
 Note that this may act as a severe constraint that oftentimes prevents the superpotential contribution in a given model. The reason is that the background $B$-field is chosen once and for all for a given model and this affects the quantisation condition on all candidate instanton divisors. The impact of the Freed-Witten anomaly on the E3-brane sector (without odd instanton fluxes) has been exemplified in \cite{Blumenhagen:2008zz,Collinucci:2008sq}.

Let us now turn to the charge of the fluxed $O(1)$ instanton under the abelian gauge groups arising from D7-branes in the compactification.  
The macroscopic $U(1)$ charge of an instanton is accounted for at a microscopic level by the appearance of charged zero modes at the intersection of the instanton with the spacetime-filling branes \cite{Blumenhagen:2006xt}. 
Let us consider a $U(1)$ instanton. The chiral index counting such charged instanton zero modes $\lambda_{EA}$ with charge $(-1_E, 1_A)$ is given by 
\bea
I_{EA} = -  \int_{X_3} [D_E] \wedge [D_A] \wedge ({  2 \pi \alpha'  ( {\cal F}_{E} - {\cal F}_{A}})),
\eea
while zero modes  $\lambda_{EA'}$ with charge $(-1_E, -1_A)$ are counted by
\bea
I_{EA'} = -  \int_{X_3} [D_E] \wedge [D_{A'}] \wedge ({2 \pi \alpha' ( {\cal F}_{E} + \sigma^*{\cal F}_{A}})).
\eea
Thus the net $U(1)_A$ charge of the zero modes is
\bea
N_A (I_{EA} - I_{EA'}) &=& \frac{1}{2} \Big( \, \kappa_{\alpha b c } \, C^{\alpha}_E  \, C^b_A \, ({\cal F}^c_A - {\cal F}^c_E) \nn \\
& & \;\; + \kappa_{\alpha \beta \gamma} C^\alpha_E \, C^\beta_A \, {\tilde{\cal F}}^\gamma_A  - \kappa_{a b \gamma} C^a_E C^b_A {\tilde{\cal F}}_E^\gamma \Big), \nn
\eea
in precise agreement with the transformation behavior (\ref{U1-trafo}) found from the supergravity analysis.

\section{Moduli stabilisation}
\label{sec:modst}

In this section we discuss moduli stabilisation for orientifold compactifications 
with $h^{1,1}_-(X_3) >0$ with special emphasis on contributions from fluxed instantons \footnote{For the case without \emph{fluxed} instantons see also \cite{Lust:2006zg}, and without gauging, but including instantons see \cite{Grimm:2007hs}.}.    Our aim 
is to highlight the differences compared to a compactification with 
$h^{1,1}_-(X_3)=0$, for which it was argued in \cite{Blumenhagen:2007sm} that it is problematic
to disentangle moduli stabilisation from the generation of a chiral open string sector.
To this end we consider an instanton-induced superpotential of the schematic form 
\bea \label{Wmod}
   W =  W_0 + \sum_{E, \cF_E} A_E({\cal F}_E)  e^{- \pi C^\alpha_{E}  T_\alpha -\tilde{q}_{E a} G^a }\ , \\
   \tilde{q}_{E a} = i\pi\kappa_{\alpha a b}  \left(C^\alpha_E \cF^b_E + C^b_E \tilde{\cF}^\alpha_E\right) \ .\nn
\eea
Note that the terms quadratic in ${\cal F}_E$ in (\ref{gaugekina}) have been absorbed into $A_E({\cal F}_E)$. This implies that
fluxed instantons with large quadratic ${\cal F}_E$ terms will 
be stronger suppressed, rendering a sum \eqref{Wmod} convergent.
The terms $W_0$, $A_E({\cal F}_E)$ are treated as constants after integrating out complex structure moduli and the dilaton, which are assumed to be stabilised at high scale using bulk background fluxes \cite{Giddings:2001yu}.

As discussed  in section \ref{sec:micu1} the divisors $D_E$ have to satisfy specific geometric conditions in order 
to support an E3-instanton entering the superpotential. Recall furthermore that the orientifold projection restricts the flux on $O(1)$ instantons to have negative orientifold parity. 
In this section we will be more general and allow also for possible superpotential contributions from $U(1)$ instantons, assuming that the additional uncharged zero modes have been saturated appropriately.
In the presence of D7-branes the sum over ${\cal F}_E$ runs only over such instanton flux configurations for which the instanton charge given in \eqref{U1-trafo} vanishes, rendering the instanton action gauge invariant. Other values of instanton flux would instead generate superpotential terms involving charged open string operators. Such instantons participate in moduli stabilisation only for non-zero VEV of the open string fields. As stressed in the introduction, at least for fields charged under the Standard Model gauge group this is not possible and will not be considered in this paper. Finally we reiterate that zero instanton charge is merely a necessary condition for the instanton to contribute as in (\ref{Wmod}). In the presence of vector-like charged zero modes with vanishing \emph{net} charge open string fields will again appear in front of $A_E$. This must be checked in concrete examples.

 It is crucial to note 
that the $G^a$ can appear in a particular combination with the K\"ahler 
moduli $T_\alpha$ as in (\ref{Wmod}), but never by themselves. Since the $G^a$ 
combine axion-like modes from the NS-NS and R-R two-form they would otherwise 
remain in the superpotential even in the
decompactification limit for $X_3$, obtained by sending all $T_{\alpha}\rightarrow \infty$. 
This is in contradiction to the fact that there is no 
potential in the ten-dimensional theory. 
In other words, if we do not include D7-branes 
which ensure that the $G^a$ are gauged as in \eqref{gauging1}, the axionic 
$\R\, G^a$ remain massless with respect to a leading order potential \eqref{Wmod}
with constant $A_E$ \cite{Grimm:2007hs}.  

Let us now consider the effect of D7-branes with fluxes. In addition to the F-term contribution arising from the superpotential above the potential will include a D-term contribution due to the gauging \eqref{gauging1} and \eqref{gauging2}. Since one is considering gauged shift symmetries with constant Killing vectors $Q_A^a,i Q_{A\alpha}$ given in \eqref{Qsodd}, \eqref{Qseven} 
the D-term is determined as $i D_A =K_{\bar G^a} Q^a_A  - i K_{\bar T_\alpha} Q_{\alpha A}$. 
Explicitly one has at leading order
\beq
\label{kahler_derivatives}
   K_{G^a} = - \frac{i}{2 \mathcal{V}}   \kappa_{a c \alpha} b^c v^\alpha\, , \qquad K_{T_\alpha}  = - \frac{v^\alpha}{ 2 \mathcal{V}}\ ,
\eeq
where $\mathcal{V}$ is the volume of $X_3$.
This yields the D-terms
\bea
  D_A &=& \frac{\ell_s^2}{2 \pi \mathcal{V}} \int_{D_A} J \wedge (\i^* B_2 - 2 \pi \alpha' \mathcal{F}_{A} )  \\
      &=& \frac{\ell_s^2}{4 \pi\mathcal{V}} v^\alpha \big(\kappa_{\alpha bc} (b^b - \mathcal{F}^b_A) C^c_A -\kappa_{\alpha \beta \gamma} \tilde \cF^\beta_A C^\gamma_A\big) \ .\nn 
\eea
Note that we have not displayed any charged matter fields which have to be added to this supergravity contribution. 

In the following we will consider compactifications to anti-de-Sitter spacetime with unbroken supersymmetry in the spirit of \cite{Kachru:2003aw}, i.e. we will require that $\left\langle F_i \right\rangle = 0 =\left\langle  D_A \right\rangle$ and $\left\langle W\right\rangle \neq 0$. Here $F_i \equiv (\partial_i + K_i) W,\ i = G^a, T_\alpha $ are the F-terms for $T_{\alpha}$, $G^a$ obtained from the superpotential \eqref{Wmod}. 

The first thing to notice is that the vanishing of the D-terms is not an independent condition. Rather it follows from the F-term equations, which read
\bea
\label{ftermT}
\frac{v^\alpha}{2\cV} = -\frac{\pi}{W} \sum_{E, \cF_E}C^\alpha_E A_E  e^{- \pi C^\alpha_{E}  T_\alpha -\tilde{q}_{E a} G^a }\ , \\
\frac{i}{2\cV}\kappa_{\alpha a b} v^\alpha b^b = -\frac{1}{W} \sum_{E, \cF_E} \tilde{q}_{E a}A_E  e^{- \pi C^\alpha_{E}  T_\alpha -\tilde{q}_{E a} G^a }.
\label{ftermG}
\eea
To see this we use the vanishing of the instanton charge \eqref{U1-trafo} to write
\be
\label{contract}
C^a_A \tilde{q}_{E a} = i\pi C^\alpha_E \left(\kappa_{\alpha a b} C^a_A \cF^b_A + \kappa_{\alpha\beta\gamma} C^\beta_A \tilde{\cF}^\gamma_A \right).
\ee
Contracting \eqref{ftermG} with $C^a_A$ and inserting \eqref{contract} together with \eqref{ftermT} then yields
$D_A=0$. This observation of course remains true in the case $h^{(1,1)}_- = 0$ considered in \cite{Blumenhagen:2007sm}. In fact, as is well-known, F- and D-term equations are never independent for vacua with non-vanishing superpotential VEV \cite{Wess:1992cp} (more recently see also e.g. \cite{Villadoro:2005}). This is due to the fact that for $\left\langle W \right\rangle \neq 0$ one can perform a K\"ahler transformation under which $K \rightarrow \tilde{K} \equiv K + \log |W|^2$. Rewriting the F- and D-terms in terms of $\tilde{K}$ it is apparent that they are proportional.

Equation \eqref{ftermT} shows that given suitable instanton wrapping numbers $C^\alpha_E$ all K\"ahler moduli can in principle be stabilised inside the K\"ahler cone. Suppose $\omega_\alpha$ is a basis of the K\"ahler cone, i.e. $J = v^\alpha \omega_\alpha$ must satisfy $v^\alpha >0$. A necessary requirement for staying in the K\"ahler cone is that every $T_\alpha$ must appear in the superpotential, i.e.  for each $\alpha = 1,\ldots , h^{(1,1)}_+$ there must be at least one \textit{uncharged} instanton contributing to \eqref{Wmod} with $C^\alpha_E \neq 0$. Note, however, that this does not require a distinct instanton to contribute to the superpotential for each $T_\alpha$ \cite{Bobkov:2010}. This general logic holds regardless of whether $h^{(1,1)}_- = 0$ or not. The advantage of a setup with $h^{(1,1)}_- > 0$ is that, as outlined in section \ref{sec:flux_Inst}, instanton fluxes act as extra degrees of freedom making the cancellation of the instanton charge due to the $T_\alpha$ generic.

Further, there are also advantages of the fluxed instanton setup from a microscopic point of view: if the net instanton charge vanishes for an unfluxed instanton due to cancellations among the charges of the $T_\alpha$ only, there remains the danger of vector-like zero modes localised on disjoint intersection curves between the instanton and the 7-branes. By contrast, in presence of disjoint intersection curves instanton flux can ensure absence of zero modes along each single curve. Concretely, for $O(1)$ instantons odd gauge fluxes ${\cal F}_A^a$ along the intersection curves with instantons can always \footnote{This is true for mutually disjoint intersection curves between the instanton and the various D7-branes; otherwise cancellation of all fluxes might be incompatible.} be canceled by opposite odd instanton flux ${\cal F}^a_E$. For $U(1)$ instantons, which may in principle allow for even instanton flux ${\cal F}^\alpha_E$, also orientifold even gauge fluxes can be compensated for in this way.  

As we noted it is not necessary for as many different instantons to appear in the superpotential as there are moduli. As an extreme example let us now demonstrate that in principle a single instanton may suffice to stabilise all the moduli with the K\"ahler moduli inside the K\"ahler cone, generalizing the work of \cite{Bobkov:2010} to the case with $h^{(1,1)}_- > 0$. For a single instanton \eqref{ftermT} shows immediately that all $v^\alpha > 0$ if and only if the instanton divisor is ample, i.e. $C^\alpha_E > 0$ $\forall \alpha$ \footnote{$C^\alpha_E < 0,$ $\forall \alpha$ is also possible, as an overall sign can be compensated by a shift of the axions $c_\alpha$.}.
Equation \eqref{ftermG} simplifies to
\be
\label{fixing_b_1inst}
i\pi \kappa_{\alpha a b} C^\alpha_E b^b = \frac{\sum_{\{ \cF_E \}} \tilde q_{E a}\tilde A_E  e^{- \tilde q_{E c} G^c }}{\sum_{\{ \cF_E \}} \tilde A_E  e^{- \tilde q_{E c} G^c }}.
\ee
Note that for a single instanton the K\"ahler moduli have dropped out and the $b^a$ can be fixed independently. The kinetic metric of the axions $c^a$ is proportional to $\kappa_{\alpha a b} v^\alpha$ and must be invertible for all $(v^\alpha)$ inside the K\"ahler cone. This shows that if the instanton is wrapped on an ample divisor all $b^a$ will be fixed by equation \eqref{fixing_b_1inst}. The success of this proposal clearly hinges upon the existence of an instanton on an ample divisor which is uncharged with respect to the gauge groups of all D7-branes present in the model. In the case with $h^{(1,1)}_- = 0$ this requires every brane flux to pull back trivially to the brane-instanton intersection. In general it will be easier to obtain an uncharged instanton on an ample divisor if $h^{(1,1)}_- > 0$, as instanton fluxes can then be used to cancel the charge. Note furthermore that the value of $\tilde q_{E a}$ is immaterial to the success of moduli stabilisation.

It is interesting to note that although the above mechanism will fix all the moduli it will generically lead to tachyonic directions in the K\"ahler moduli. This follows from a result in \cite{Conlon:2006tq} where it was shown that in a supersymmetric AdS vacuum each unfixed axion has a tachyonic superpartner if only the F-terms are considered. Although the superpartner is tachyonic it is still Breitenlohner-Freedman stable in AdS. We would then require that the uplift mechanism can keep these tachyons stable after uplifting to Minkowski or de-Sitter space \footnote{For example, an uplift potential of the form $V = \frac{C}{\cV^2}$ as considered in \cite{Kachru:2003aw} leads to a positive definite contribution to the mass matrix of the K\"ahler moduli and therefore can ensure absence of tachyons in the de-Sitter case.}. The presence of the $U(1)$s modifies this only in that if an axion is eaten by a $U(1)$ its superpartner will receive a positive mass contribution from the D-term rendering it non-tachyonic. Therefore altogether the number of tachyons is given by the number of moduli minus the number of linearly independent instantons in the superpotential minus the number of massive $U(1)$s. 

To summarize, the effect on moduli stabilisation of the addition of D7-branes in a compactification with $\left\langle F_i \right\rangle = 0 =\left\langle  D_A \right\rangle$ and $\left\langle W\right\rangle \neq 0$ along the lines of \cite{Kachru:2003aw} is to reduce the number of uncharged instantons which may contribute to the superpotential. This problem can be ameliorated by considering compactifications with $h^{(1,1)}_- > 0$ in which instanton fluxes can cancel the instanton charge. The D-terms associated to the brane gauge groups do not give rise to additional constraints in this type of compactification.

The latter does not generally hold in the case of compactifications to Minkowski space or with broken supersymmetry. 
The moduli stabilisation versus chirality problem in  non-supersymmetric vacua was considered in concrete examples with $h^{1,1}_- =0$ in \cite{Blumenhagen:2007sm,Collinucci:2008sq,Blumenhagen:2009gk} with the conclusion that in the models with no unstabilised moduli, the D-term drives a subset of the moduli to the boundary of the K\"ahler cone.
It would be important to determine whether this is the result of a general no-go theorem or not.


\section{Charged instantons in F-theory}
\label{sec:instf}

In this section we comment on two independent effects relevant for moduli stabilisation in F-theory \footnote{For recent work on instantons in F-theory see e.g. \cite{Blumenhagen:2010ja}. The four-dimensional F-theory effective action for the harmonic modes
has been derived via M-theory in \cite{Grimm:2010ks}.}.

The first issue concerns the role of massive abelian gauge symmetries and their selection rules in F-theory. 
It is key to carefully distinguish the flux-induced St\"uckelberg mechanism associated with the  gauging  \eqref{Qseven}  of the K\"ahler moduli $T_\alpha$ on the one hand and the geometric St\"uckelberg mechanism due to
the gauging \eqref{Qsodd}  of the $G^a$ moduli on the other hand. The latter can only occur for $h^{1,1}_-(X_3) \neq 0$ and is independent of 7-brane gauge flux.

$U(1)$ gauge bosons which become massive via the geometric St\"uckelberg mechanism in Type IIB language turn out to receive masses at the Kaluza-Klein scale in F-theory and are absent in the low-energy regime. As was suggested in \cite{Grimm:2010ez} and is further worked out in the upcoming \cite{appear} such geometrically massive $U(1)$ potentials can be consistently described by non-harmonic forms in supergravity. The absence of the abelian symmetry at low energies raises the question of the role of D-terms and selection rules for instantons in F/M-theory. 

Let us first turn on no gauge fluxes for the geometrically massive $U(1)$s. The $U(1)$ symmetry and the corresponding D-terms are integrated out at the KK-scale. The uplift of the B-field moduli are fixed at their D-flat value $b^a=0$ at the KK-scale. The axionic partners become longitudinal modes of the massive $U(1)$ bosons, which are described by expansion of $C_3$ into non-harmonic forms. 
The Type IIB instanton expression can indeed be matched explicitly with the M5-instanton partition function \cite{appear}. What remains is the analogue of the F-terms \eqref{ftermT} for those M5-instantons that contribute to the superpotential.

Being integrated out at low energies, the massive abelian symmetries do not act as \emph{manifest} selection rules in supergravity, but they do remain as \emph{accidental} symmetries of the low-energy theory. As such they have exactly the same effect as in the Type IIB weak coupling limit.
In particular, the same selection rules act on M5-instantons in F-theory as on Type IIB E3-brane instantons. This is because the  analogue of the charged instanton zero modes at the intersection with the 7-branes are still present even if the $U(1)$ has a mass at the KK scale. Furthermore they may in addition carry charge under the non-abelian gauge group of the 7-branes and can thus not be neglected. The same conclusions as for IIB regarding moduli stabilisation and chirality follow and the same proposal of using instanton fluxes will also be relevant.

If we do switch on gauge fluxes associated with the geometrically massive $U(1)$s we must take into account the non-harmonic forms describing the massive gauge bosons as well  \cite{appear}. This is because in M-theory such flux translates into $G_4$-fluxes that are expanded into non-harmonic forms. In this case the D-terms cannot be integrated out at high energies, and the picture for solving the F- and D-terms resembles the procedure in the Type IIB weak coupling limit.

To summarise, if we consider the F-theory uplift of a Type IIB orientifold compactification, the role of $U(1)$ selection rules and their impact on moduli stabilisation is effectively unchanged in F-theory.
By contrast, not every F-theory compactification is smoothly connected to a Type IIB orientifold. For those F-theory models which do not have a smooth Type IIB weak coupling limit, \emph{extra} effects associated with exceptional gauge symmetries do modify and, in fact, improve the situation of moduli stabilisation, as we discuss next.


The reason why, following \cite{Blumenhagen:2007sm}, we have not included instanton contributions in \eqref{Wmod} involving charged open string fields is because
any open-string modes charged under the $U(1)$ part of the D7 gauge group must also be charged under the non-abelian part  in Type IIB orientifold compactifications. 
Since the non-abelian part must remain unbroken - at least for the Standard Model gauge group - such open fields cannot acquire a VEV. This excludes the dressed instanton from the set relevant for moduli stabilisation.
The new ingredient, which is intrinsic to F-theory, is the presence of open string modes which are charged under the $U(1)$ supporting the chirality generating flux (thereby rendering some instanton charged), but which are not charged under the non-abelian part of the visible gauge group. 

To see the presence of such modes it is useful to adopt the approach advocated in \cite{Beasley:2008dc} and describe the intersection of F-theory 7-branes by the localised enhancement of the gauge group. 
For example in an $SU(5)$ GUT a point intersection between three 7-branes corresponds to a rank 2 enhancement of the gauge group to $SO(12)$, $SU(7)$ or $E_6$, and the localised modes are extracted by decomposing the adjoint of the enhanced gauge group. The first two cases correspond to intersections that can arise in IIB while the latter is intrinsic to F-theory. Decomposing the adjoint we have for instance
\bea
E_6 &\supset& SU(5) \times U(1)_{a'} \times U(1)_{b'} \;, \\
\bf{78} &\rightarrow& \bf{24}^{(0,0)} \op \un^{(0,0)} \op \un^{(0,0)} \op \\ \nn
&&  [ \un^{(-5,-3)} \op \f ^{(-3,3)} \op \te ^{(-1,-3)}  \op \te ^{(4,0)} + c.c. ]. \nn
  \label{fe6}
\eea
The $SU(5)$ singlets charged under the $U(1)$s are the appropriate fields to dress the charged instantons and make them gauge invariant. There is no general phenomenological constraint which forbids a vacuum expectation value for these singlets. Thus the appropriate instanton contribution can be present. A non-zero VEV for such a singlet Higgses the $U(1)$ and removes the associated selection rule. From a microscopic perspective the Higgsing lifts the charged zero modes. This can enable the instanton to contribute to the moduli stabilisation superpotential. 
Of course as always it must be checked whether the mass matrix in the instanton effective action is really of sufficient rank to lift \emph{all} charged zero modes.

Despite this caveat, the genuinely F-theoretic feature is that the Higgsing of the $U(1)$ is compatible with an unbroken Standard Model gauge group. 
 Indeed it is simple to check that there are no charged $SU(5)$ (or more generally Standard Model) singlets for  Type IIB models as these are ultimately based on the gauge group $SO(2N)$. Put differently, in F-theory models based on exceptional gauge symmetry, chirality is compatible with the Higgsing of all abelian $U(1)$ symmetries. In fact, the Higgsed phase is the generic one in moduli space. 


{\noindent \bf Acknowledgments}

We thank R. Blumenhagen, A. Hebecker and C. Mayrhofer for discussions. T.W. acknowledges hospitality of the MPI, Munich. T.G. thanks the Simons Center at Stony Brook, Cornell University and MIT for hospitality and support. This work was supported in part by the DFG under TR33 "The Dark Universe" and by the European ERC Advanced Grant 226371 MassTeV and the PITN contract PITN-GA-2009-237920.

%
%

\end{document}